# Thermal Boundary Conductance Across Metal-Nonmetal Interfaces: Effects of Electron-Phonon Coupling both in Metal and at Interface

Mengjie Li[1], Yuanyuan Wang[2], Jun Zhou[1,a], Jie Ren[2,b], and Baowen Li[1,4,5]

[1] Center for Phononics and Thermal Energy Science, School of Physics Science and Engineering, Tongji University, Shanghai 200092, People's Republic of China

[2] School of Urban Development and Environmental Engineering, Shanghai Second Polytechnic University, Shanghai 201209, People's Republic of China

[3] Department of Chemistry, Massachusetts Institute of Technology, Cambridge, MA 02139, USA

[4] Department of Physics, Center for Computational Science and Engineering, and Graphene Research Center, National University of Singapore, Singapore 117546, Republic of Singapore

[5] NUS Graduate School for Integrative Sciences and Engineering, National University of Singapore, Singapore 117456, Republic of Singapore

**Abstract.** We theoretically investigate the thermal boundary conductance across metal-nonmetal interfaces in the presence of the electron-phonon coupling not only in metal but also at interface. The thermal energy can be transferred from metal to nonmetal via three channels: (1) the phonon-phonon coupling at interface; (2) the electron-phonon coupling at interface; and (3) the electron-phonon coupling within metal and then subsequently the phonon-phonon coupling at interface. We find that these three channels can be described by an equivalent series-parallel thermal resistor network, based on which we derive out the analytic expression of the thermal

---

a  e-mail：zhoujunzhou@tongji.edu.cn

b  e-mail:jieustc@gmail.com

boundary conductance. We then exemplify different contributions from each channel to the thermal boundary conductance in three typical interfaces: Pb-diamond, Ti-diamond, and TiN-MgO. Our results reveal that the competition among above channels determines the thermal boundary conductance.

# 1 Introduction

The thermal boundary conductance (TBC), which is the capability of heat conduction across an interface between two dissimilar materials, plays an important role in the design of electronics devices [1,2]. The investigation on the TBC across metal-nonmetal interfaces, which is the total heat current $Q$ divided by the temperature drop at the interface shown in Figure 1(a), is one of the most important topics for thermal engineering. Experimentally, the TBC across metal-nonmetal interfaces is measured with the thermoreflectance technique [3-7] and the steady state technique [8]. One of the concerns to researchers is that some experimentally measured values significantly deviate from the theoretical calculated ones [9,10] where only phonons are considered. The methods for the TBC calculations that only consider phonon transport include the acoustic mismatch model (AMM), the diffuse mismatch model (DMM) [11], and the lattice dynamical method (LDM) [12]. Some of the calculated results from these methods overestimate the TBC while some calculated results underestimate the TBC. The reason is that the phonons are the major heat carriers in semiconductors and insulators while the electrons are the major heat carriers in metals [13]. Therefore, the contributions from both electrons and phonons to the TBC across metal-nonmetal interfaces must be considered by introducing three channels of heat conductions. The heat conduction via the phonon-phonon (PP) coupling between phonons in metal and phonons in nonmetal is noted as Channel (1). The heat conduction via the electron-phonon (EP) coupling between electrons in metal and phonons in nonmetal is noted as Channel (2). The heat conduction via the EP coupling within metal followed by a subsequent PP coupling at interface is noted as Channel (3). Figure 1(b) shows the schematic diagram of these three channels: the red arrows represent Channel (1) with heat current $Q_1$; the yellow arrows represent Channel (2) with heat current $Q_2$; and the blue arrows represent channel (3) with heat current $Q_3$. The thermal transport processes in Figure 1(b) can be described by a thermal resistor network as shown in Figure 1(c) which includes the interfacial PP resistance $R_{pp}$, the interfacial EP resistance $R_{ep}$, the volumetric electronic thermal resistance $\delta x/\kappa_m^e$, the volumetric lattice thermal resistance $\delta x/\kappa_m^p$, and the volumetric EP resistance $1/(\delta x G)$ due to the EP coupling. Here, we divide the interfacial area into slabs with infinitesimal width $\delta x$. $\kappa_m^e$ ($\kappa_m^p$) is the electronic (lattice) thermal conductivity and $G$ is the EP coupling constant in metals.

The contribution from Channel (2) to the TBC has been studied individually by many researchers. Theoretically, Reich [14] and Zhang *et al.* [15]. studied the TBC due to the EP coupling at interface in one dimensional chain model. Hopkins and Norris [16] investigated the contributions from electron-interface scatterings to the TBC using the relaxation time approximation. Huberman and Overhauser [17] calculated the TBC across Pb-diamond interface due to the EP coupling by considering the joint vibrational modes near the interface. The TBC they calculated is in agreement with the value measured by references [9,10]. Sergeev [18] treated the EP coupling at interface in analogous to the inelastic electron-impurity scattering and found that the TBC is proportional to the inverse of EP energy relaxation time. Mahan [19] calculated the TBC by considering the EP coupling at interface due to image potential generated by ion charges. Ren and Zhu [20] found an asymmetric and negative differential TBC by considering nanoscale metal-nonmetal interfaces. From experimental approach, Hopkins *et al.* [21] measured the EP coupling constant in gold films. The substrate dependency and film thickness dependency of the EP coupling constant were observed, which proves the importance of the EP coupling at interface.

The contribution from Channel (3) to the TBC has also been studied separately, by excluding other two channels, from both theoretical and experimental approaches. Majumdar and Reddy [22] found that the PP interfacial resistance and the volumetric EP resistance in metal are in series. The TBC across TiN-MgO interface they calculated is in agreement with the experimental data measured by Costescu *et al.* [7] Taking a step forward, Singh *et al.* [23] studied the TBC through detailed consideration of the EP coupling constant from the Bloch-Boltzmann-Perierls formula. Battayal *et al.* [8] measured and calculated the TBC across Al-diamond interface. Their results show that the volumetric EP resistance and the interfacial PP resistance are in the same order of magnitude.

However, there are not many works that consider three channels simultaneously and compare the contributions from different channels. Li *et al.* [24] measured the thermal conduction in periodic Mo-Si multilayers using $3\omega$ method. They adopted all three channels to interpret the measured results. Basu *et al.* [13] measured the thermal conductivity in metal-semiconductor nanocomposites. A reduction of thermal conductivity is interpreted as the enhanced PP scattering at interfaces between nanoparticles and the presence of the EP coupling both inside metallic nanocrystals and at interfaces [25]. In references [13,24], all three channels are taken into account to explain their experimental results. Nevertheless, there is no general theoretical model to calculate the TBC when all channels are considered.

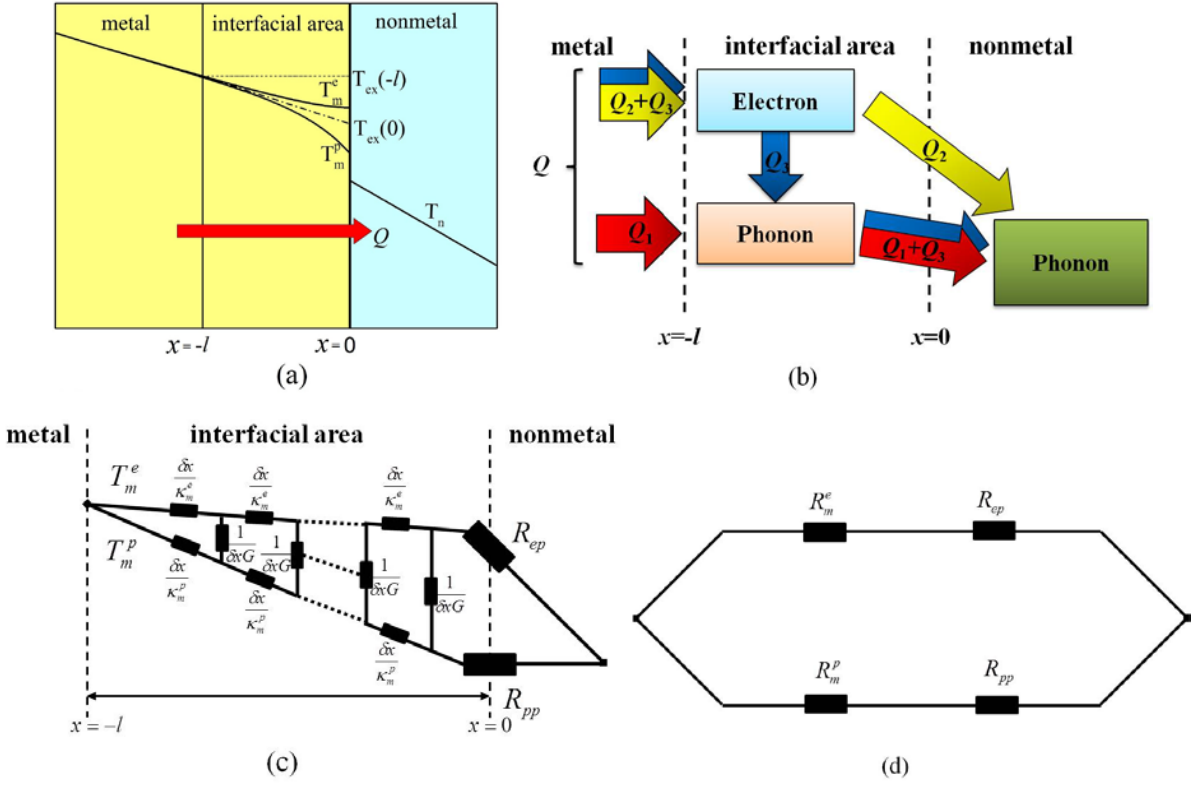

Fig. 1. (color online) (a) Schematic illustration of heat conduction from metal to nonmetal across the interface along $x$-direction. $Q$ is the total heat current and $l$ is the width of interfacial area, defined in the main text. (b) Schematic of three channels of heat conduction: Channel (1) is in red; Channel (2) is in yellow; Channel (3) is in blue. The total heat current consists of the heat current of every channel, $Q = Q_1 + Q_2 + Q_3$. (c) Complicated thermal resistor network which includes all three channels. (d) Equivalent series-parallel thermal resistor network that gives a same TBC as (c), where $R_m^e = l/\kappa_m^e$ ($R_m^p = l/\kappa_m^p$) is the renormalized electronic (lattice) thermal resistance.

In this paper, we use the two-temperature model (TTM) of heat conduction to calculate the TBC by considering all three channels mentioned above simultaneously. It is shown that the TBC of metal-nonmetal systems can be interpreted as a series-parallel thermal resistor network. Our results could be useful for interpreting the role of the EP coupling both in metal and at interface on the TBC. We further explain why some measured TBC is lower than the calculated value while some measured TBC is higher than the calculated one. The paper is organized as follows: Section 2 presents the modified model of TBC using TTM by considering all three channels we mentioned above. In Section 3, we apply our model to three typical metal-nonmetal interfaces: Pb-diamond, Ti-diamond, and TiN-MgO to exemplify the reason of the difference between theoretical and experimental. Section 4 concludes the paper.

## 2 Model

We consider the heat conduction along the *x*-direction as shown in Figure 1(a). We do the analytical calculation by using the TTM [26], of which predictions had been validated by the experimentally measured nonequilibrium transport of electrons and phonons using short-pulsed laser excitations [27]. In the TTM, the electron (phonon) subsystem in metal is characterized by the electron (phonon) temperature $T_m^e$ ($T_m^p$) and the temperature in nonmetal is $T_n$. The energy transfer between the electron subsystem and the phonon subsystem in metal is proportional to their temperature difference $G(T_m^e - T_m^p)$ [28,29]. Using the Fourier's law, the energy balance equations for steady state in the absence of energy source can be written as [26,30]:

$$\kappa_m^e \frac{d^2 T_m^e(x)}{dx^2} - G[T_m^e(x) - T_m^p(x)] = 0, \tag{1a}$$

$$k_m^p \frac{d^2 T_m^p(x)}{dx^2} + G[T_m^e(x) - T_m^p(x)] = 0, \tag{1b}$$

$$k_n \frac{d^2 T_n(x)}{dx^2} = 0. \tag{1c}$$

Here $k_n$ is the lattice thermal conductivity of the nonmetal. It is obvious that $T_m^e = T_m^p$ deep inside the metal which means that electron and phonon subsystems are in equilibrium far away from the interface. We use $l = (G/\kappa_m^e + G/\kappa_m^p)^{-1/2}$ to define the width of interfacial area as shown in Figure 1 [22]. Clearly, in the interfacial area, the electron temperature and the phonon temperature are never equilibrium with each other even in the steady state. And the $l$ characterizes the decay length of the temperature discrepancy, i.e. $T_m^e(x) - T_m^p(x) \sim \exp[-|x|/l]$. Solving above equations, $T_m^e$, $T_m^p$, and $T_n$ can be analytically obtained in the forms of:

$$T_m^e(x) = \alpha e^{x/l} + \beta x + \xi, \tag{2a}$$

$$T_m^p(x) = -\alpha e^{x/l} + \beta x + \xi, \tag{2b}$$

$$T_n(x) = \varepsilon x + \delta, \tag{2c}$$

where the coefficients need to be determined. Both $T_m^e$ and $T_m^p$ asymptotically approach the extrapolation of the linear temperature profile $T_{ex}(x) = \beta x + \xi$ as shown in Figure 1(a). All the coefficients in equation (2) can be solved by using the boundary conditions due to the heat flux continuity at interface

$$-\kappa_m^e \frac{dT_m^e(x)}{dx}\bigg|_{x=0} = [T_m^e(0) - T_n(0)]/R_{ep}, \tag{3a}$$

$$-\kappa_m^p \frac{dT_m^p(x)}{dx}\bigg|_{x=0} = [T_m^p(0) - T_n(0)]/R_{pp}, \tag{3b}$$

$$-\kappa_m^e \frac{dT_m^e(x)}{dx}\bigg|_{x=0} -\kappa_m^p \frac{dT_m^p(x)}{dx}\bigg|_{x=0} = -\kappa_n \frac{dT_n(x)}{dx}\bigg|_{x=0}. \tag{3c}$$

After obtain the temperature profiles, we calculate the TBC using the following definition

$$\sigma_K = \frac{Q}{T_{ex}(-l) - T_n(0)}. \tag{4}$$

Here we use the temperature drop between the extrapolation of the linear temperature profile $T_{ex}(-l)$ and the temperature $T_n(0)$ as shown in Figure 1(a). This definition is different from $Q/[T_{ex}(0)-T_{n}(0)]$, which uses the temperature drop between $T_{ex}(0)$ and $T_{n}(0)$ [22]. In the physical sense of heat conduction, the area within $-l<x<0$ should be an important part of the interface since the nonequilibrium between electrons and phonons strongly affects the transport properties within this area. Then the analytical expression of the overall TBC, as:

$$\sigma_K = \frac{1}{R_m^e + R_{ep}} + \frac{1}{R_m^p + R_{pp}}, \qquad (5)$$

Here $R_m^e = l/\kappa_m^e$ is the renormalized electronic thermal resistance and $R_m^p = l/\kappa_m^p$ is the renormalized lattice thermal resistance. Equation (5) describes the TBC exactly the same as a series-parallel thermal resistor network as depicted in Figure 1(d): $R_m^e$ connects to $R_{ep}$ in series which is Channel (3); $R_m^p$ connects to $R_{pp}$ in series which is the combination of Channel (2) and Channel (3); and then these two series resistances are connected in parallel. Therefore, the complicated thermal transport network in Figure 1(c) can be treated as an equivalent series-parallel thermal resistor network in Figure 1(d).

In the limit case of $\kappa_m^e \gg \kappa_m^p$ in most metals, the width of interfacial area is about $l \approx \sqrt{\kappa_m^p/G}$; the renormalized electronic thermal resistance is negligible $R_m^e \approx 0$; and the renormalized lattice thermal resistance can be approximated as $R_m^p \approx 1/\sqrt{G\kappa_m^p}$. As such, equation (5) can be approximated as

$$\sigma_K \approx \frac{1}{R_{ep}} + \frac{1}{R_m^p + R_{pp}}. \qquad (6)$$

It means that $R_{pp}$ and $R_m^p$ are firstly in series and then in parallel with $R_{ep}$. The second term on the right side of equation (6) recovers Majumdar and Reddy's results in reference [22]. From equation (6), it is easy to find that $\sigma_K < 1/R_{pp}$ when $1/R_{pp} > 1/(2R_{ep}) + \sqrt{1/(2R_{ep})^2 + 1/(R_m^p R_{ep})}$ and $\sigma_K > 1/R_{pp}$ for the other cases. This feature provides a possible reason that measured TBC ($\sigma_K^{\exp}$) does not agree with the calculated values ($\sigma_K^{\text{theo}}$) that only considers the phonon transport, in other words, approximating $\sigma_K^{\text{theo}}$ by $1/R_{pp}$. For example, $\sigma_K^{\exp} < \sigma_K^{\text{theo}}$ for interface such as TiN-MgO, and $\sigma_K^{\exp} > \sigma_K^{\text{theo}}$ for interface like Pb-diamond and Ti-diamond. In order to eliminating the disagreement between $\sigma_K^{\exp}$ and $\sigma_K^{\text{theo}}$, both the interfacial EP resistance $R_{ep}$ and the renormalized lattice thermal resistance $R_m^p$, must be considered together with interfacial PP resistance $R_{pp}$ as shown in equations (5) and (6).

## 3 Results and discussions

In the following, we exemplify the contributions of $R_{ep}$, $R_m^e$, and $R_m^p$ to the TBC in three typical metal-nonmetal interfaces we mentioned above: Pb-diamond, Ti-diamond, and TiN-MgO.

**Pb-diamond interface**

We first studied the TBC across Pb-diamond interface. The measured TBC $\sigma_K^{\exp}$ at temperature 273K is $0.031\text{GW}/(\text{m}^2 \cdot \text{K})$ [9] and the calculated TBC $\sigma_K^{\text{theo}} = 1/R_{pp}$ using the DDM that only considers phonon transport is $0.002\text{GW}/(\text{m}^2 \cdot \text{K})$ [9], which is much smaller than the measured one. In our calculation, we choose the EP coupling constant in Pb to be $G = 1.24 \times 10^{17}\text{W}/(\text{m}^3 \cdot \text{K})$ [31]. The electronic thermal conductivity is estimated as $\kappa_m^e = 33.6\text{W}/(\text{m} \cdot \text{K})$ by using the Wiedemann-Franz law. $\kappa_m^p = 2.4\text{W}/(\text{m} \cdot \text{K})$ is estimated by subtracting $\kappa_m^e$ from the total thermal conductivity which is $36 \text{W}/(\text{m} \cdot \text{K})$ [32]. The width of interfacial area is estimated as $l \approx 4.25 \text{nm}$. The renormalized electronic and lattice thermal resistances are $R_m^e \approx 0.126 [\text{GW}/(\text{m}^2 \cdot \text{K})]^{-1}$ and $R_m^p \approx 1.77 [\text{GW}/(\text{m}^2 \cdot \text{K})]^{-1}$, respectively. Using above parameters, in Figure 2(a) we show $1/R_{ep}$ as a function of $1/R_{pp}$ when fixing $\sigma_K$ to be the experimentally measured value $0.031 \text{GW}/(\text{m}^2 \cdot \text{K})$. The calculated interfacial PP resistance from DMM and LDM is $R_{pp} \approx 500 [\text{GW}/(\text{m}^2 \cdot \text{K})]^{-1}$ [9]. Then we find that $1/R_{ep}$ should be about $0.03 \text{GW}/(\text{m}^2 \cdot \text{K})$, which means $R_{ep} \approx 33 [\text{GW}/(\text{m}^2 \cdot \text{K})]^{-1}$. Therefore, we conclude that $\sigma_K \sim 1/R_{ep}$ since $R_{ep} \gg R_m^e$ and $R_{ep} + R_m^e \ll R_{pp} + R_m^p$. This result testifies that Channel (2) dominates the TBC across Pb-diamond interface as shown in the inset of Figure 2(a). Our conclusion is consistent with reference [17].

**Ti-diamond**

For the Ti-diamond interface, the measured TBC $\sigma_K^{\exp}$ at temperature 293K is $0.1 \text{GW}/(\text{m}^2 \cdot \text{K})$ [9] and the calculated TBC $\sigma_K^{\text{theo}} = 1/R_{pp}$ using the LDM that only considers phonon transport is about $0.065 \text{GW}/(\text{m}^2 \cdot \text{K})$ [9]. The EP coupling constant in Ti is $G = 1.3 \times 10^{18} \text{W}/(\text{m}^3 \cdot \text{K})$ [33]. The electronic thermal conductivity is estimated as $\kappa_m^e = 16.7 \text{W}/(\text{m} \cdot \text{K})$ by using the Wiedemann-Franz law. $\kappa_m^p = 5.6 \text{W}/(\text{m} \cdot \text{K})$ is estimated by subtracting $\kappa_m^e$ from the total thermal conductivity which is $22.3 \text{W}/(\text{m} \cdot \text{K})$ [34]. Then, the width of interfacial area is $l \approx 1.8 \text{nm}$. The renormalized electronic and lattice thermal resistances are $R_m^e \approx 0.11[\text{GW}/(\text{m}^2 \cdot \text{K})]^{-1}$ and $R_m^p \approx 0.32[\text{GW}/(\text{m}^2 \cdot \text{K})]^{-1}$, respectively. Based on the above parameters, in Figure 2(b) we show $1/R_{ep}$ as a function of $1/R_{pp}$ when fixing $\sigma_K$ to the experimentally measured value $0.1\text{GW}/(\text{m}^2 \cdot \text{K})$. We adopt the value of interfacial PP resistance around $R_{pp} \approx 15.4[\text{GW}/(\text{m}^2 \cdot \text{K})]^{-1}$ that is calculated from LDM [9]. Then the corresponding $1/R_{ep}$ should be about $0.036\text{GW}/(\text{m}^2 \cdot \text{K})$ which means $R_{ep} \approx 27[\text{GW}/(\text{m}^2 \cdot \text{K})]^{-1}$. We thus conclude that $\sigma_K \sim 1/R_{ep} + 1/R_{pp}$ since $R_{ep}$ and $R_{pp}$ are on the same order and both of them are much larger than $R_m^e$ and $R_m^p$. Therefore, for Ti-diamond interface, the renormalized electronic and lattice thermal resistances can be ignored. The interfacial EP resistance and the interfacial PP resistance are connected in parallel and are equally important, in other words, both Channels (1) and (2) need to be considered as shown in the inset of Figure 2(b).

**TiN-MgO**

For TiN-MgO interface, the measured TBC $\sigma_K^{\text{exp}}$ at temperature 293K is $0.72\text{GW}/(\text{m}^2 \cdot \text{K})$ [7] and the calculated TBC $\sigma_K^{\text{theo}} = 1/R_{pp}$ using the DDM that only considers phonon transport is about $1.03\text{GW}/(\text{m}^2 \cdot \text{K})$ [22]. The EP coupling constant in TiN is estimated as $G \sim 2.6 \times 10^{18} \text{W}/(\text{m}^3 \cdot \text{K})$, which is much larger than the estimated value in reference [22]. Here we used the expression $G = 3\hbar\gamma\lambda <\omega^2>/(\pi k_B)$ [29], where $\hbar$ is the Planck constant, $k_B$ is the Boltzmann constant, $\gamma \sim 211\text{J}/(\text{m}^3 \cdot \text{K}^2)$ [35] is the electronic specific heat constant, $\lambda \sim 0.59$ is EP coupling parameter [36] and $<\omega^2>$ is the mean square frequencies of phonons with $\hbar <\omega^2>^{1/2} \sim 30\text{meV}$ [37]. The total thermal conductivity in TiN is about $28.84\text{W}/(\text{m}\cdot\text{K})$ [38]. The electronic thermal conductivity in TiN can hardly be estimated since the Wiedemann-Franz law is not valid in this material. Therefore, we study cases with three different $\eta = \kappa_m^e/\kappa_m^p$ with fixed $\kappa_m^e + \kappa_m^p$. The width of interfacial area can then be estimated as $l \approx 0.65\text{nm}$ when $\eta = 24$, $l \approx 0.71\text{nm}$ when $\eta = 20$, and $l \approx 0.8\text{nm}$ when $\eta = 15$, respectively. The renormalized electronic and lattice thermal resistances are $R_m^e \approx 0.023[\text{GW}/(\text{m}^2\cdot\text{K})]^{-1}$ and $R_m^p \approx 0.56[\text{GW}/(\text{m}^2\cdot\text{K})]^{-1}$ for $\eta = 24$, $R_m^e \approx 0.026[\text{GW}/(\text{m}^2\cdot\text{K})]^{-1}$ and $R_m^p \approx 0.52[\text{GW}/(\text{m}^2\cdot\text{K})]^{-1}$ for $\eta = 20$, $R_m^e \approx 0.03[\text{GW}/(\text{m}^2\cdot\text{K})]^{-1}$ and $R_m^p \approx 0.44[\text{GW}/(\text{m}^2\cdot\text{K})]^{-1}$ for $\eta = 15$, respectively. Using above parameters, in Figure 2(c) we show $1/R_{ep}$ as a function of $1/R_{pp}$ when $\sigma_K$ is fixed to the experimentally measured value $0.72\text{GW}/(\text{m}^2\cdot\text{K})$ [7]. We adopt the value of $1/R_{pp}$ around $1.03\text{GW}/(\text{m}^2\cdot\text{K})$ that is calculated from DMM and LDM [22], which means $R_{pp} \approx 0.97[\text{GW}/(\text{m}^2\cdot\text{K})]^{-1}$. We find that $1/R_{ep}$ is smaller than $0.1\text{GW}/(\text{m}^2\cdot\text{K})$, which means $R_{ep} > 10[\text{GW}/(\text{m}^2\cdot\text{K})]^{-1}$, for all reasonable values of $\eta$ we studied. Therefore, we conclude that $\sigma_K \sim 1/(R_m^p + R_{pp})$ since $R_{ep} + R_m^e \gg R_{pp} + R_m^p$ and $R_m^p \sim R_{pp}$. This result testifies the fact that in TiN-MgO interface, both Channels (1) and (3) dominate the TBC as shown in the inset of Figure 2(c).

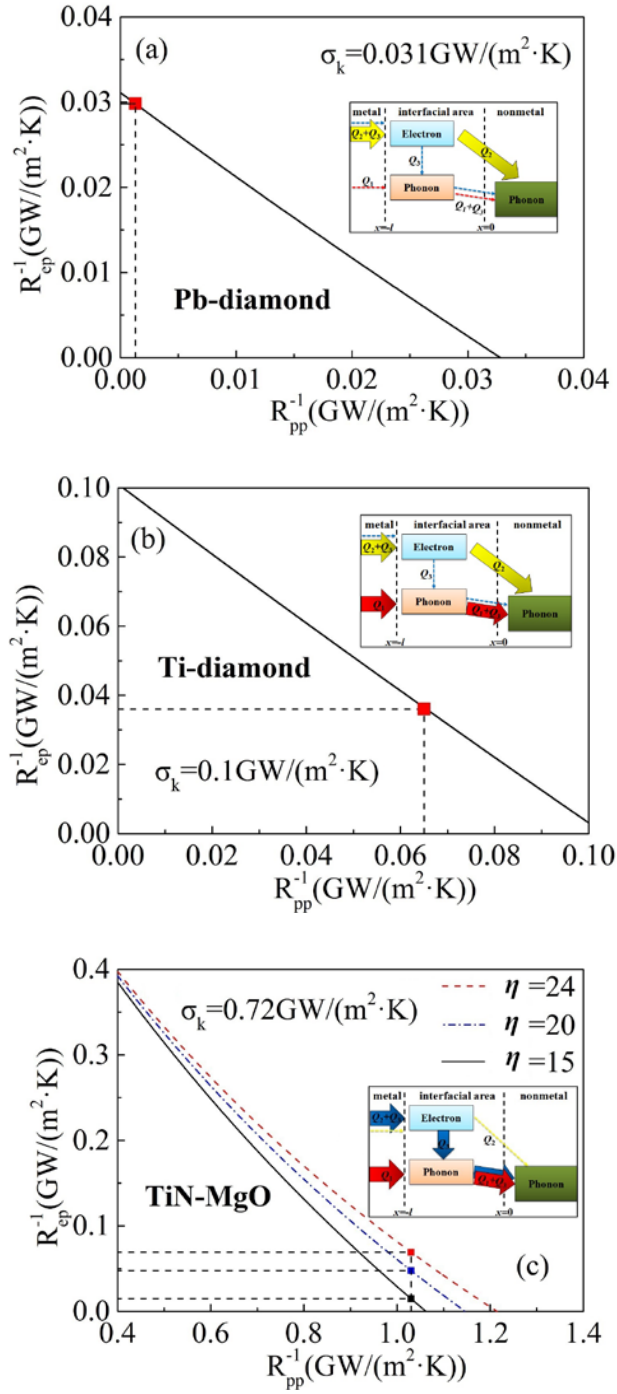

Fig. 2. (color online) $1/R_{ep}$ as a function of $1/R_{pp}$ for (a) Pb-diamond, (b) Ti-diamond, and (c) TiN-MgO interfaces when $\sigma_K$ is fixed to be the experimental measured values. The vertical

dashed line is the calculated $1/R_{pp}$ from the DMM and the LDM in reference [7,9]. The insets show the relative importance among the three channels. The dashed arrows are the negligible channels.

## 4 Conclusion

We have given an analytical expression for the thermal boundary conductance across a metal-nonmetal interface with considering the electron-phonon couplings both in metal and at interface. The calculation is achieved under the two-temperature model of heat conduction by extending the concept of interface thermal resistance with the contribution from an interfacial area. We have shown that the thermal boundary conductance can be model as a series-parallel thermal resistor network which considers three heat conduction channels, including interfacial electron-phonon resistance $R_{ep}$, renormalized electronic thermal resistance $R_m^e$, renormalized lattice thermal resistance $R_m^p$, and interfacial phonon-phonon resistance $R_{pp}$. Our model successfully explains the observations that: i) in Pb-diamond interface, $\sigma_K \sim 1/R_{ep}$, and in Ti-diamond interface $\sigma_K \sim 1/R_{pp} + 1/R_{ep}$; the ignorance of $R_{ep}$ gives the reason that the theoretical TBC only considering phonon transport underestimates the measured values; ii) in some interfaces like TiN-MgO, $\sigma_K \sim 1/(R_{pp} + R_m^p)$; the ignorance of $R_m^p$ gives the reason that the theoretical TBC only considering phonon transport overestimates the measured values.

**Acknowledgements**

This work is supported by the National Natural Science Foundation of China (Grant No. 11334007). J. Z. is also supported by the program for New Century Excellent Talents in Universities (Grant No. NCET-13-0431) and the Program for Professor of Special Appointment (Eastern Scholar) at Shanghai Institutions of Higher Learning (Grant No. TP2014012). Y. W. is also supported by the Shanghai Natural Science Foundation (Grant No. 14ZR1417000). J. R. acknowledges the hospitality of the Center for Phononics and Thermal Energy Science, Tongji University, where this work was carried out.

**Reference:**
1. P. L. Kapitza, J. Phys. USSR **4**, 181 (1941).
2. D. G. Cahill, *et al.*, J. Appl. Phys. **93**, 793 (2003).
3. R. J. Stevens, A. N. Smith, and P. M. Norris, J. Heat Trans. **127**, 315 (2005).
4. H. Lyeo and D. G. Cahill, Phys. Rev. B **73**, 144301 (2006).
5. L. Guo, *et al.*, J. Heat Trans. **134**, 042402 (2012).
6. P. E. Hopkins, *et al.*, Int. J. Thermophys. **28**, 947 (2007).
7. R. M. Costescu, M. A. Wall, and D. G. Cahill, Phys. Rev. B **67**, 054302 (2003).
8. M. Battayal, *et al.*, Diamond Relat. Mater. **17**, 1438 (2008).
9. R. J. Stoner and H. J. Maris, Phys. Rev. B **48**, 16373 (1993).
10. R. J. Stoner, *et al.*, Phys. Rev. Lett. **68**, 1563 (1992).
11. E. T. Swartz and R. O. Pohl, Rev. Mod. Phys. **61**, 605 (1989).
12. D. A. Young and H. J. Maris, Phys. Rev. B **40**, 368 (1989).
13. T. S. Basu, *et al.*, Appl. Phys. Lett. **103**, 083115 (2013).
14. K. V. Reich, Prog. Theor. Exp. Phys. 03I01 (2013).
15. L. Zhang, J. Lü, J. Wang, and B. Li, J. Phys.:Condens Matter **25**, 445801 (2013).
16. P. E. Hopkins and P. M. Norris, J. Heat Transfer **131**, 043208 (2009).
17. M. L. Huberman and A. W. Overhauser, Phys. Rev. B **50**, 2865 (1994).
18. A. V. Sergeev, Phys. Rev. B **58**, 10199(R) (1998).
19. G. D. Mahan, Phys. Rev. B 79, 075408 (2009).
20. J. Ren and J. X. Zhu, Phys. Rev. B **87**, 231412(R) (2013).
21. P. E. Hopkins, J. L. Kassebaum, and P. M. Norris, J. Appl. Phys. **105**, 023710 (2009).
22. A. Majumdar and P. Reddy, Appl. Phys. Lett. **84**, 4768 (2004).
23. P. Singh, M. Seong, and S. Singh, Appl. Phys. Lett. **102**, 181906 (2013).
24. Z. Li, *et al.*, Nano Lett. **12**, 3121 (2012).
25. J. Ordonez-Miranda, *et al.*, J. Appl. phys. **109**, 094310 (2011).

26. T. Q. Qiu and C. L. Tien, ASME J. Heat Transfer **115**, 835 (2993).

27. J. Fujimoto, *et al.*, Phys. Rev. Lett. **53**, 1837 (1984).

28. M. I. Kaganov, I. M. Lifshitz, and M. V. Tanatarov, Sov. Phys. JETP. 4,173 (1957).

29. P. B. Allen, Phys. Rev. Lett. **59**, 1460 (1987).

30. S. I. Anisimov, B. L. Kapeliovich, and T. L. Perel'man, Sov. Phys. JETP **39**, 375 (1974).

31. M. A. Al-Nimr, Int. J. Thermophysics **18**, 5(1997).

32. U. Mizutani, *Introduction to the Electron Theory of Metals* (Cambridge University Press, Cambridge, 1995).

33. Z. Lin, L. V. Zhigilei, and V. Celli, Phys. Rev. B **77**, 075133 (2008).

34. V. E. Peletskii, High Temp. -High Pressures **17**, 111 (1985).

35. W. Weber, Phys. Rev. B **8**, 5093 (1973).

36. E. I. Isaev, *et al.*, J. Appl. Phys. **101**, 123519 (2007).

37. W. Spengler, *et al.*, Phys. Rev. B **17**, 1095 (1978).

38. W. Lengauer, *et al.*, J. Alloys Comp. **217**, 137 (1995).